\documentclass[lettersize,journal]{IEEEtran}
\usepackage{amsmath,amsfonts,amssymb,amsthm}
\usepackage{xcolor}
\usepackage[dvipsnames]{xcolor}
\usepackage{algorithm}
\usepackage{algpseudocode}
\usepackage{array}
\usepackage{textcomp}
\usepackage{stfloats}
\usepackage{url}
\usepackage{verbatim}
\usepackage{graphicx}
\usepackage{cite}
\usepackage{multirow}
\usepackage{caption}
\usepackage{subcaption}
\usepackage{hhline}

\def\BibTeX{{\rm B\kern-.05em{\sc i\kern-.025em b}\kern-.08em
    T\kern-.1667em\lower.7ex\hbox{E}\kern-.125emX}}
\usepackage{balance}

\def\b0{{\pmb{0}}} 

\begin{document}
	
\title{Evolution of UE in Massive MIMO Systems \\
for 6G: From Passive to Active}

 \author{
Kwonyeol Park, Hyuckjin Choi, Geonho Han, Gyoseung Lee, Yeonjoon Choi, Sunwoo Park, and Junil Choi
}



\maketitle

\begin{abstract}
As wireless networks continue to evolve, stringent latency and reliability requirements and highly dynamic channels expose fundamental limitations of gNB-centric massive multiple-input multiple-output (mMIMO) architectures, motivating a rethinking of the user equipment (UE) role. In response, the UE is transitioning from a passive transceiver into an active entity that directly contributes to system-level performance.
In this context, this article examines the evolving role of the UE in mMIMO systems during the transition from fifth-generation (5G) to sixth-generation (6G), bridging third generation partnership project (3GPP) standardization, device implementation, and architectural innovation. 
Through a chronological review of 3GPP Releases 15 to 19, we highlight the progression of UE functionalities from basic channel state information (CSI) reporting to artificial intelligence (AI) and machine learning (ML)-based CSI enhancement and UE-initiated beam management. We further examine key implementation challenges, including multi-panel UE (MPUE) architectures, on-device intelligent processing, and energy-efficient operation, and then discuss corresponding architectural innovations under practical constraints.  
Using digital-twin-based evaluations, we validate the impact of emerging UE-centric functionalities, illustrating that UE-initiated beam reporting improves throughput in realistic mobility scenarios, while a multi-panel architecture enhances link robustness compared with a single-panel UE.

\end{abstract}


\section{Introduction}
Massive multiple-input multiple-output (mMIMO) has been a cornerstone technology in fifth-generation (5G) wireless communication systems, particularly within 5G new radio (NR) \cite{larsson2014massive}. By leveraging large-scale antenna arrays deployed at the gNodeB (gNB), mMIMO enhances spectral efficiency and maximizes system capacity, thereby enabling the high data rates required for enhanced mobile broadband (eMBB). These capabilities have underpinned the successful deployment of 5G networks.

The transition toward sixth-generation (6G) wireless communication systems, however, marks a paradigm shift that extends far beyond throughput improvements. 
Emerging services, such as extended reality (XR), integrated sensing and communication (ISAC), and non-terrestrial networks (NTN), demand unprecedented levels of latency, reliability, and ubiquitous connectivity. Meeting these requirements necessitates a fundamental evolution of the mMIMO framework.

Against this backdrop, prior works on mMIMO have largely focused on optimizing precoding strategies at the gNB, analyzing theoretical capacity gains from scaling antenna arrays, and refining gNB-side signal processing \cite{liu2023nr}. In contrast, the role of user equipment (UE), which directly affects end-to-end system performance, has received relatively limited attention. 
In 6G systems, the UE is no longer a passive device that merely executes gNB-configured commands. Instead, it is evolving into an intelligent entity capable of sensing, interpreting, and proactively responding to environmental dynamics while contributing directly to system-level optimization. 
This transition reshapes the mMIMO design paradigm and motivates a more systematic examination of UE-centric evolution. 

This article provides an integrated analysis of the evolving role of the UE in mMIMO systems by bridging three dimensions: its evolution in the third generation partnership project (3GPP) standardization, practical hardware implementation, and future architectural innovations. We posit that the UE must be redefined as an intelligent node equipped with on-device processing capabilities, multi-panel UE (MPUE) architectures, and energy-efficient designs, rather than serving solely as an endpoint for connectivity. The main contributions are summarized as follows.

\begin{itemize}
    \item \textbf{Evolution in 3GPP standardization:} We review the evolution of UE capabilities from Release~15 to Release~19, highlighting the shift from a passive role limited to channel state information (CSI) reporting to active functionalities in Releases~18 and~19, including artificial intelligence (AI) and machine learning (ML)-based CSI enhancement and UE-initiated beam management.
    \item \textbf{UE implementation challenges:} We examine key device-level challenges, including MPUE architectures to overcome form-factor and blockage constraints, on-device intelligence to reduce signaling overhead, and energy-efficient operation under dynamic traffic and channel conditions.
    \item \textbf{Structural innovations:} Beyond single-device optimization, we discuss architectural innovations to address system-level limitations, including capability-based UE classification, collaborative UE networking, and AI-enabled hardware platforms.
    \item \textbf{Digital-twin-based validation:} We validate emerging UE-centric functionalities using high-fidelity digital twin simulations, demonstrating throughput and robustness gains from UE-initiated beam reporting and MPUE architectures.
\end{itemize}

Collectively, these discussions highlight the central role of the UE in shaping the performance and robustness of future mMIMO systems.

\section{Evolution of the UE Role in 3GPP}
\label{Chap2}
The evolution of the UE within 3GPP standardization, spanning Releases~15 to~19, reflects a transition from passive reporting to active intelligence. During the initial 5G NR deployment, the UE primarily served as a passive information provider, measuring and reporting CSI for gNB scheduling and beam management. As channel dimensionality and signaling overhead increased, more sophisticated UE functionalities became necessary. By Releases~18 and~19, the UE has evolved into an intelligent entity capable of AI/ML-enabled processing and proactive beam reporting \cite{Lin_2025_3GPPRel19Evolution}.

\subsection{Release~15: Foundation of mMIMO and Passive UE} 
\label{Sec2-1}

3GPP Release~15 established the foundational mMIMO framework for 5G NR, primarily targeting beamforming gain and spatial multiplexing through large-scale gNB antenna arrays \cite{ghosh20195g}. In this initial architecture, the UE supported basic beam discovery by measuring synchronization signal blocks (SSBs) transmitted by the gNB over predefined beam directions, mainly for initial access and coarse alignment. This procedure enabled the UE to identify suitable transmission beams but did not involve fine-grained link optimization.

Beyond this limited role, the primary function of the UE was to acquire and report channel state information (CSI) for resource allocation, link adaptation, and precoder selection at the gNB. Specifically, the UE measured channel state information reference signals (CSI-RSs) and fed back parameters such as the precoding matrix indicator (PMI), rank indicator (RI), and channel quality indicator (CQI) \cite{3gpp.38.213, 3gpp.38.214}. In time division duplex (TDD) systems, uplink–downlink reciprocity partially reduced this feedback burden via sounding reference signal (SRS)-based channel estimation \cite{3gpp.38.211}.

Despite this foundation, Release~15 faced limitations, particularly in frequency division duplex (FDD) systems where explicit CSI feedback scaled poorly with the number of gNB antenna ports. Moreover, the passive nature of UE measurement and reporting procedures often resulted in beam misalignment under high-mobility scenarios, motivating enhancements in subsequent releases.

\subsection{Release~16 / 17: Enhancing Efficiency and Robustness}
\label{Sec2-2}
Releases~16 and~17 focused on improving efficiency and robustness in practical deployments. A major enhancement was the reduction of CSI feedback overhead in FDD systems. Release~16 introduced the enhanced Type-II codebook, leveraging frequency-domain correlation and spatial-domain compression to preserve high-resolution channel information while minimizing UE feedback overhead. Building upon this advancement, Release~17 incorporated angle-delay reciprocity to infer downlink CSI from uplink SRS measurements.

TDD systems were also optimized to enhance uplink MIMO capability and overall transmission efficiency. Release~16 enabled flexible antenna switching, allowing the UE to transmit SRS with different antenna configurations, while Release~17 extended support for up to eight receive antenna chains, significantly enhancing UE-side reception performance \cite{jin2023massive}. In addition, the partial SRS frequency hopping improved resource utilization by enabling efficient frequency sharing among multiple UEs.

To address high-mobility scenarios, Release~17 introduced the unified transmission configuration indicator (TCI) framework. By harmonizing beam indications for control and data channels, this framework minimized signaling overhead and beam switching latency, enabling more reliable beam tracking under rapid channel variations.

\subsection{Release~18 / 19: Intelligent UE with AI/ML Integration}
\label{Sec2-3}
With the advent of 5G-Advanced in Releases~18 and~19, mMIMO has evolved from a performance-driven architecture into an intelligent system, redefining the UE as an active and intelligent entity. This shift has been driven by diverse UE form factors, highly dynamic channel conditions, and the integration of AI/ML technologies.

Release~18 focused on uplink enhancement and mobility resilience. Motivated by advanced UEs such as fixed wireless access (FWA) devices and automotive terminals with more flexible antenna configurations, Release~18 introduced support for up to eight uplink layers and expanded demodulation reference signal (DMRS) ports, with a maximum of 24 ports, to boost multi-user MIMO capacity \cite{jin2023massive}. To combat channel aging in high-mobility scenarios, the UE in Release~18 actively predicts upcoming channel conditions by exploiting Doppler characteristics extracted from received reference signals. 

In addition, AI/ML became a core component of the CSI framework. Release~18 initiated studies on AI/ML-based CSI reporting, and Release~19 advanced this effort by standardizing a two-sided model for CSI prediction and compression, allowing the UE to learn complex channel features locally \cite{Lin_2025_3GPPRel19Evolution}. Furthermore, Release~19 introduced UE-initiated beam management \cite{R12504472}, empowering the UE to autonomously detect beam degradation and trigger updates. This capability ensures robust connectivity under rapid channel variations, moving beyond reliance solely on network-driven control.

\section{Practical Challenges in UE Implementation}
\label{Chap3}
While 3GPP standardization has established functional specifications, a substantial gap remains between the performance envisioned in these specifications and the quality of experience (QoE) delivered in commercial deployments. As 6G demands unprecedented throughput, latency, and reliability, implementation challenges are no longer confined to the gNB but place significant demands on the UE.

Unlike the gNB, the UE faces stringent limitations in form factor, battery life, thermal dissipation, and cost. Consequently, the implementation of standardized functionalities on commercial UEs demands advancements in hardware architecture and intelligent control mechanisms that operate efficiently under these constraints. This section examines three core implementation challenges from the UE perspective: MPUE architectures, on-device intelligent processing, and energy-efficient operation. 

\begin{figure}[!t]
    \centering    \includegraphics[width=0.5\textwidth]{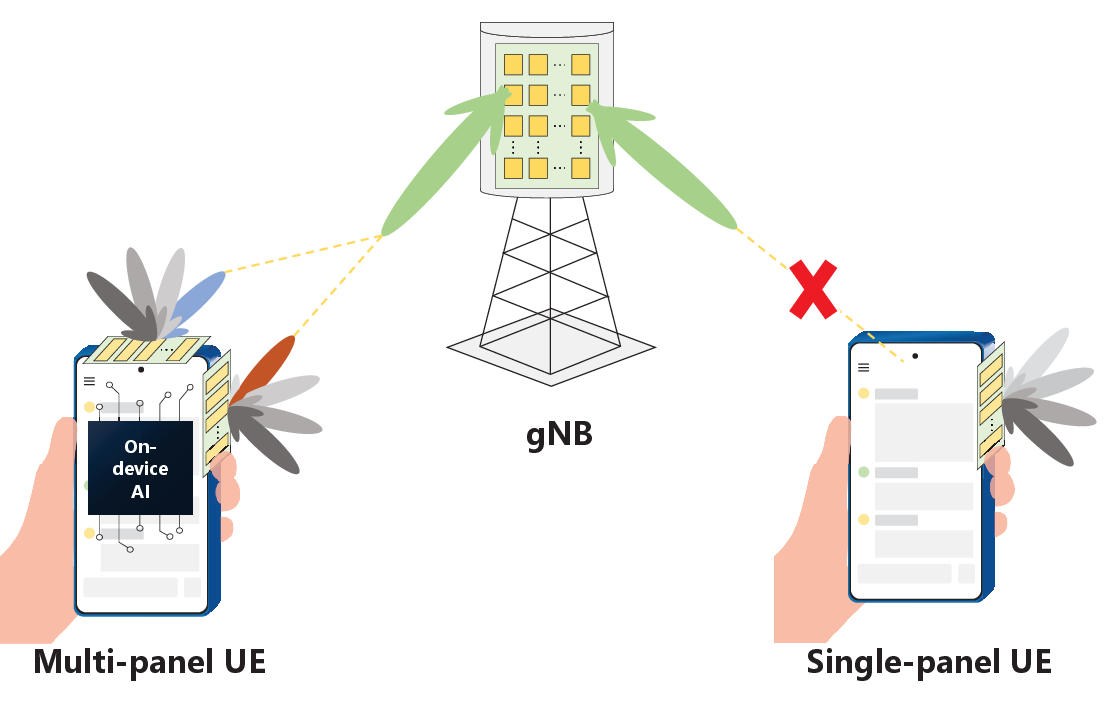}
    \caption{Conceptual comparison of link robustness between multi-panel UE and single-panel UE.}
    \label{fig:MPUE}
\end{figure}

\subsection{Multi-Panel UE Architecture}
\label{Sec3-1}
As mMIMO systems continue to develop, the number of antennas deployed at the gNB has increased steadily, whereas the UE has inherent limitations on scaling up its antenna array due to strict form factor constraints. To tackle this asymmetry and mitigate the severe path loss experienced in millimeter-wave (mmWave) and frequency range 3 (FR3) bands, the MPUE architecture has emerged as a key enabler \cite{heng2021six}. Rather than relying on a single antenna panel, the MPUE architecture distributes multiple compact antenna panels across different regions of the device to enhance spatial coverage and link robustness, as illustrated in Fig.~\ref{fig:MPUE}.

The practical implementation of MPUE architectures involves significant hardware complexity. To efficiently utilize the limited space, compact antenna panels can be placed on the front, back, and side edges of the UE by adopting an edge-oriented design. However, such an architecture increases the number of radio frequency (RF) chains, leading to higher power consumption and circuit design complexity. Also, at high-frequency bands, hand-induced signal blockage becomes a dominant performance-limiting factor, making optimized panel layout essential to avoid blockage or exploit hand-reflection gains.

Beyond physical layout considerations, intelligent panel control algorithms are essential to cope with dynamic channel conditions. During initial access or handover procedures, activating multiple panels can improve measurement accuracy by maximizing the spatial information captured. To reduce unnecessary power consumption, context-aware panel switching algorithms proactively select an optimal combination of panels by considering environmental contexts such as user grip patterns and mobility as well as instantaneous signal strengths. Furthermore, cross-layer designs that tightly couple physical-layer beam management with higher-layer resource scheduling are critical for maintaining service continuity and stable performance in MPUE-based systems. The effectiveness of the MPUE architecture is quantitatively evaluated in Section~\ref{Chap5}.

\subsection{On-Device Intelligent Processing Capabilities}
\label{Sec3-2}
Advanced AI/ML functionalities introduced in 3GPP Releases~18 and~19 impose new implementation requirements on the UE, particularly under the ultra-low-latency and high-reliability demands of immersive services such as extended reality (XR) and vehicle-to-everything (V2X). In these scenarios, decision-making based solely on network or cloud-centric processing is often insufficient due to signaling delay and limited reactivity to fast channel variations. As a result, intelligent processing capabilities are increasingly pushed toward the UE, enabling local inference and control. Such on-device intelligence improves responsiveness while reducing signaling overhead and computational burden on the network in mMIMO systems.

Beam tracking and prediction rely heavily on AI-based models to proactively mitigate beam misalignment. Notably, the UE-initiated beam reporting mechanism standardized in Release~19 enables the UE to trigger reports only when needed by using on-device intelligence to anticipate beam-quality degradation, thereby reducing signaling overhead and reaction time. In addition, on-device AI enables efficient CSI compression by distilling high-dimensional channel information into compact representations, reducing uplink feedback overhead and gNB-side processing.

Beyond these functions, on-device intelligence enhances context awareness for mMIMO systems by exploiting locally available UE information. By leveraging device-specific data such as location, mobility state, and sensor measurements, the UE can provide the gNB with valuable side information. Moreover, the UE can detect interference patterns and either apply proactive mitigation locally or report refined interference-related metrics to the gNB. This capability complements gNB-centric interference management and improves the accuracy of system-level link adaptation. Furthermore, integration with federated learning and digital-twin frameworks enables environment-adaptive operation, improving the scalability and robustness of mMIMO deployments.

\subsection{Energy-Efficient UE Operation}
\label{Sec3-3}
In mMIMO systems, the UE must sustain substantial real-time computation to support multi-panel operation, continuous beam tracking, and high-resolution CSI reporting, inevitably increasing power consumption. Given the limited battery capacity of the UE, enabling low-power operation while preserving battery lifetime is a prerequisite for the practical deployment of advanced UE capabilities, not merely a performance optimization.

Commercial data traffic exhibits strong spatiotemporal variability. Under these conditions, operating the baseband at a fixed voltage and clock frequency is energy-inefficient. Accordingly, the UE baseband should incorporate an intelligent power-management framework that dynamically adapts its operating state to the instantaneous computational workload and quality of service (QoS) requirements. Implementing such a framework requires coordinated integration of multiple low-power techniques. Specifically, dynamic voltage and frequency scaling (DVFS) adjusts the supply voltage and clock frequency in real time based on computational demand \cite{mo2018energy}. In parallel, fine-grained power and clock gating selectively disable inactive beamforming paths or antenna panels to suppress leakage power. In addition, sleep control mechanisms such as quick sleep and discontinuous reception (DRX) exploit traffic burstiness by activating high-speed links only during data-intensive periods and rapidly transitioning into deep-sleep states during micro-idle intervals.

In summary, UE baseband architectures for 6G mMIMO systems must evolve into intelligent power-management platforms with tightly integrated hardware and software. Such platforms are essential to jointly deliver high throughput and ultra-low latency under stringent battery constraints, thereby enabling sustainable and scalable deployment of advanced UE capabilities.

\section{Structural Innovations for 6G UE}
\label{Chap4}
Building upon the implementation challenges discussed earlier, this section explores the structural innovations toward 6G mMIMO systems, with a focus on UE-centric architectures. In this context, the UE is no longer defined by a single specification but is differentiated according to service requirements and expected to evolve from an isolated device into an active participant in collaborative network architectures.

\begin{table*}[!t]
\centering
\caption{Three UE types and capability configurations in 3GPP discussions \cite{R12505414}}
\label{tab:UE_types}
\renewcommand{\arraystretch}{1.1}
\setlength{\tabcolsep}{4.5pt}
\begin{tabular}{|c|l|l|l|l|}
\hline
\textbf{} & \textbf{Representative device and service} & \textbf{Peak data rate} & \textbf{Maximum bandwidth} & \textbf{Maximum MIMO layer} \\
\hline
\textbf{Type-A} & High-performance eMBB (Smartphone, XR, FWA, CPE) 
& \begin{tabular}[c]{@{}l@{}}DL: 10~Gbps\\ UL: 2~Gbps\end{tabular}
& \begin{tabular}[c]{@{}l@{}}DL: $\geq$ 200~MHz\\ UL: 100 / 200~MHz\end{tabular}
& \begin{tabular}[c]{@{}l@{}}DL: $\geq$ 4\\ UL: $\geq$ 2\end{tabular} \\
\hline
\textbf{Type-B} & Mid-tier eMBB (Wearable, XR-lite) 
& \begin{tabular}[c]{@{}l@{}}DL: 200~Mbps-1~Gbps \\ UL: 50~Mbps-200~Mbps\end{tabular}
& \begin{tabular}[c]{@{}l@{}}DL: 100~MHz\\ UL: 100~MHz\end{tabular}
& \begin{tabular}[c]{@{}l@{}}DL: 1-2\\ UL: 1\end{tabular} \\
\hline
\textbf{Type-C} & Low-power IoT (Sensor, Industrial device)
& \begin{tabular}[c]{@{}l@{}}DL: 10~Mbps\\ UL: 5~Mbps\end{tabular}
& \begin{tabular}[c]{@{}l@{}}DL/UL: 5~MHz for FDD \\ DL/UL: 20~MHz for TDD\end{tabular}
& \begin{tabular}[c]{@{}l@{}}DL: 1-2 \\ UL: 1\end{tabular} \\
\hline
\end{tabular}
\end{table*}
\subsection{Redefining UE Types for 6G}
\label{Sec4-1}
In 6G systems, the wireless ecosystem is expected to support a wide range of UEs, from ultra-high-throughput XR devices to ultra-low-power IoT sensors and automotive platforms \cite{amiri2024application}. This diversity challenges the conventional one-size-fits-all UE design philosophy, which becomes impractical under highly heterogeneous service requirements. Recent 3GPP discussions have therefore shifted toward a modular UE classification framework in which UEs are categorized by capability levels and scenarios. These categories are commonly denoted as Type-A, Type-B, and Type-C, as summarized in Table.~\ref{tab:UE_types}.

Type-A UEs represent the highest-capability category, encompassing XR terminals, FWA devices, and high-end smartphones that demand extreme data rates, such as up to 10~Gbps in the downlink and 2~Gbps in the uplink. To support wide bandwidths and multiple MIMO layers, often exceeding four downlink layers and two uplink layers, these devices must integrate a large number of antennas and RF chains, which inevitably increases hardware complexity. Beyond serving as high-performance transceivers, Type-A UEs are also expected to function as active control entities. They can leverage powerful on-device processing to autonomously execute key physical-layer functions such as channel prediction, beam tracking, and interference management. By optimizing radio conditions locally through embedded AI, they can sustain ultra-low-latency service quality while reducing network-side signaling overhead.

In contrast to Type-A UEs with extreme performance targets, Type-B UEs correspond to mid-capability devices such as XR-lite smart glasses and wearable terminals, for which mobility and form-factor constraints are dominant design drivers. Although their performance requirements are lower than those of Type-A UEs, they still demand substantially higher connectivity than Type-C UEs. The key challenge for Type-B UEs is to sustain robust mMIMO performance under tight constraints on size, thermal dissipation, and power. Rather than increasing the number of physical antennas, Type-B UEs depend on efficiency-oriented techniques, including intelligent beam management and lightweight AI models, to maintain stable connectivity while tightly controlling power consumption.

At the other end of the spectrum, Type-C UEs include low-power devices such as industrial sensors and compact IoT nodes, for which battery lifetime and cost efficiency are top priorities. For these devices, mMIMO is leveraged not to maximize data rates but to improve spatial isolation and energy efficiency. By exploiting high-resolution beamforming at the gNB, mutual interference among a massive number of devices can be effectively suppressed and separated, thereby reducing retransmissions and improving reliability. In turn, Type-C UEs can increase sleep durations and significantly extend operational lifetime.

\subsection{UE-Centric mMIMO for Collaborative Networking}
\label{Sec4-2}
The physical form-factor constraints of individual UEs are expected to become major bottlenecks for mMIMO performance. To address these limitations, a UE-centric mMIMO paradigm has emerged, in which multiple nearby devices, such as smartphones, wearables, and IoT sensors, are virtually aggregated and coordinated to operate as a single distributed large-scale antenna array. This approach increases cooperative diversity and spatial rank, thereby substantially expanding communication capacity. In addition, the enlarged effective aperture formed by distributed UEs enables centimeter-level localization accuracy, creating new opportunities for high-precision positioning services.

To enable distributed devices to operate as a coherent virtual array, real-time data exchange among cooperating UEs is essential. Conventional Layer-2 or Layer-3 solutions, including tethering and Wi-Fi connectivity, incur millisecond-level latency due to protocol overhead, rendering them unsuitable for latency-sensitive services such as XR. Therefore, ultra-low-latency data relaying at the physical layer is required, where signals are forwarded directly without higher-layer protocol processing. Moreover, cooperation across heterogeneous frequency bands, such as simultaneous operation in FR2 and FR3, may lead to numerology misalignment among devices. Addressing this issue requires advanced synchronization techniques and interoperable standards for rapid inter-frequency-range conversion.

Forming a virtual antenna array further requires accurate knowledge of the relative geometry among cooperating UEs, including inter-device distances, orientations, and antenna layouts. This motivates three-dimensional relative positioning based on time-of-arrival and angular measurements exchanged among devices, complemented by hybrid localization that incorporates local sensor data such as inertial measurement unit (IMU) readings. When devices operate in close proximity, the far-field assumption breaks down and spherical-wave propagation must be considered. In such near-field regimes, accurate signal modeling is essential to mitigate phase errors and enable coherent joint transmission and reception.

Beyond communication, the extended aperture enabled by UE-centric mMIMO unlocks advanced sensing capabilities, particularly for UE-centric ISAC \cite{ISAC_intro_1}. For example, a set of cooperating vehicles or pedestrian devices can jointly form a distributed MIMO radar, enabling environmental sensing with fewer blind spots and higher spatial resolution. Such cooperative sensing, however, requires exchanging context-sensitive information tightly coupled to user location, mobility, and surrounding context, which raises significant privacy concerns. Addressing these issues calls for integrated privacy-preserving mechanisms based on on-device processing and federated learning, ensuring that sensitive raw data remain local while enabling collaborative intelligence across the network.

\subsection{AI/ML Hardware Accelerators and Federated Learning}
\label{Sec4-3}
To practically realize the intelligent processing capabilities discussed earlier, including on-device AI processing and the UE-centric collaboration, fundamental innovations in UE hardware architecture are required. Conventional designs that rely solely on general-purpose processors are insufficient to support the massive matrix computations and real-time AI inference workloads inherent to mMIMO systems, particularly under stringent power and latency constraints.

To meet this demand, the integration of dedicated AI accelerators, such as neural processing units (NPUs), into UE modem architectures is expected to increase. Such specialized hardware can be optimized for the complex-valued matrix operations characteristic of mMIMO processing, delivering orders-of-magnitude gains in energy efficiency, often quantified in tera operations per second (TOPS) per watt, compared with general-purpose processors. This enables high-performance AI inference to be executed locally at the UE within tight power budgets and transmission time intervals on the order of milliseconds.

In addition to raw computational capability, a major bottleneck for on-device AI stems from data movement within the UE rather than computation itself. Channel measurements from large antenna arrays must be transferred repeatedly between memory and processing units, which incurs substantial latency and power consumption. To mitigate this overhead, emerging architectural techniques such as processing-in-memory (PIM) and intelligent caching are essential. By performing computation within or near memory, these approaches reduce data movement, thereby lowering latency and improving system-level energy efficiency.

Finally, extending UE intelligence from individual devices to the network level calls for federated learning platforms. Each UE trains AI models locally using its own environment-specific observations, while only model updates, such as gradients or parameter deltas, are shared with the infrastructure. This approach preserves user privacy by keeping raw data on-device, while still enabling global models that capture diverse propagation conditions. As a result, a hybrid processing architecture emerges in which real-time inference is performed locally at the UE, and large-scale training and aggregation are handled by the network. Under this paradigm, the UE becomes an autonomous learning node that continuously adapts within the 6G ecosystem.

\begin{figure}[!t]
    \centering    \includegraphics[width=0.5\textwidth]{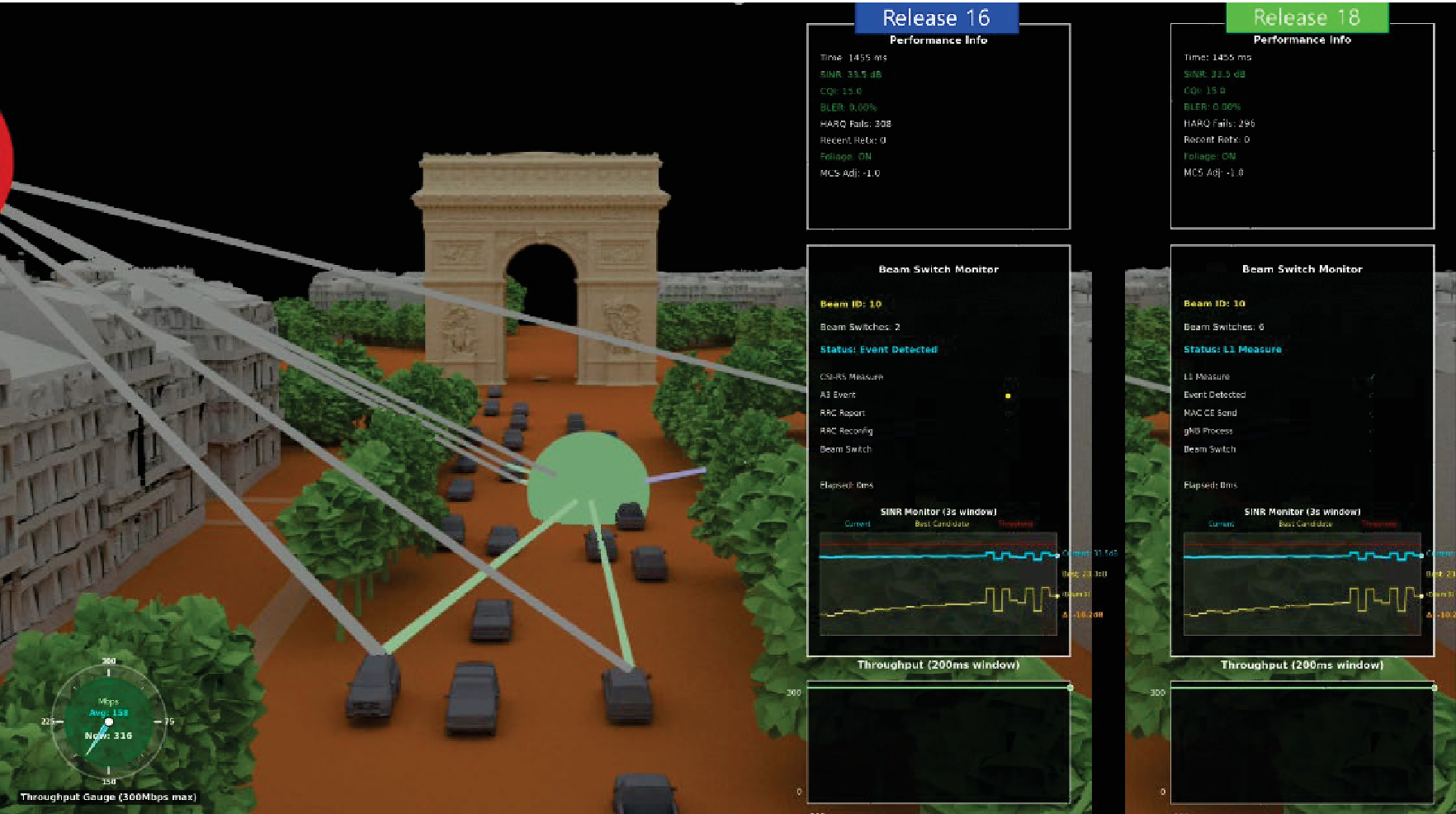}
    \caption{Digital twin framework for mMIMO experiments in Paris, powered by NVIDIA Sionna RT.}
    \label{fig:sionna}
\end{figure}

\section{Simulation Results and Analysis}
\label{Chap5}
This section validates the UE-centric design principles discussed earlier using high-fidelity digital-twin simulations. We quantitatively evaluate the impact of UE-initiated beam reporting introduced in 3GPP Release~19 and MPUE architectures on system throughput and signal-to-interference-plus-noise ratio (SINR) under realistic mobility and propagation conditions.

\subsection{Role of Digital Twin and Simulation Setup}
\label{Sec5-1}
As 6G mMIMO systems grow in complexity, evaluation needs to capture realistic propagation, mobility, and beam-management behaviors, beyond abstract channel models or isolated link-level analysis. Digital twin platforms provide a unified, high-fidelity environment that can replicate real deployment conditions in a controlled manner, enabling systematic system-level evaluation under practical operating scenarios.

Our simulations are conducted using the NVIDIA Sionna framework as a digital twin platform \cite{hoydis2023sionna}. We emulate a dense urban deployment scenario by incorporating detailed three-dimensional city geometry and realistic mobility patterns, as shown in Fig.~\ref{fig:sionna}. Specifically, vehicular UEs move at a speed of 20~km/h, mimicking realistic city traffic with frequent blockages and line-of-sight (LoS) to non-LoS transitions. Channel propagation is generated using the built-in ray-tracing (RT) engine in NVIDIA Sionna, which computes geometry-consistent multipath components along UE trajectories within the 3D scene. This setup enables accurate modeling of beam dynamics between a gNB and vehicular UEs, allowing comprehensive evaluation of beam management performance under realistic propagation and mobility conditions.


\begin{figure}[!t]
    \centering    \includegraphics[width=.5\textwidth]{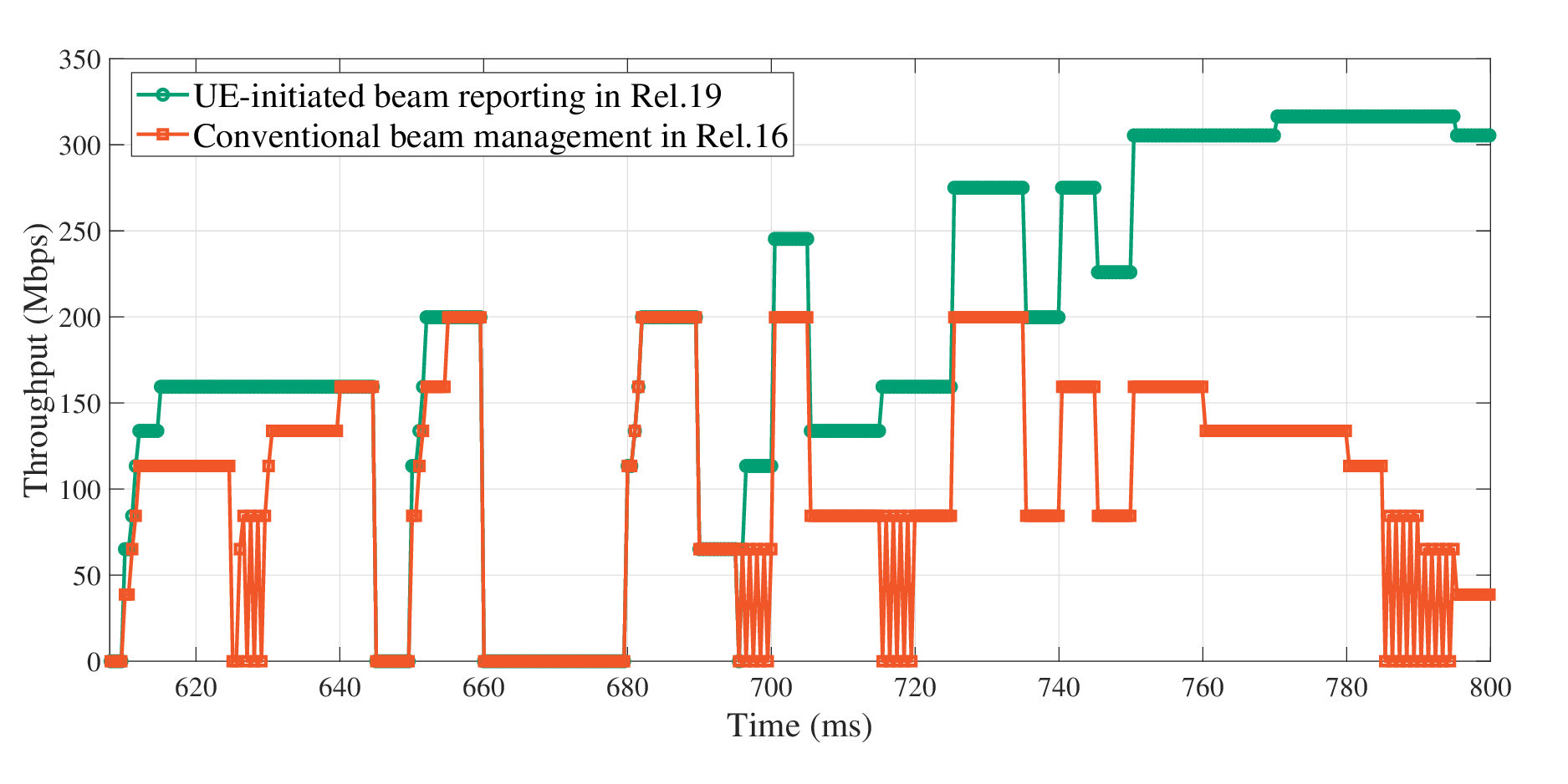}
    \caption{Throughput comparison between Release~16 and Release~19 in a mobility scenario.}
    \label{fig:tput_comparison}
\end{figure}

\subsection{Performance of UE-Initiated Beam Reporting}
\label{Sec5-2}
In realistic mobility scenarios, dynamic channel variations make timely beam alignment particularly challenging. Under the legacy gNB-centric scheme specified in Release~16, beam management is largely driven by higher-layer configuration. The UE performs Layer-1 measurements on configured SSB/CSI-RS resources and reports them according to radio resource control (RRC)-defined procedures, which are often periodic in practice. This introduces a temporal gap between beam-quality degradation and the subsequent beam update, increasing the likelihood of beam misalignment.

To evaluate this effect, we implement the Release~19 UE-initiated beam reporting mechanism and compare it against the Release~16 baseline. In the Release~19 mechanism, the UE can trigger beam reporting based on instantaneous Layer-1 measurements and configured triggering conditions, enabling more timely beam updates through lower-layer signaling. Note that explicit prediction or on-device intelligence is not employed in this implementation, and the reporting trigger is purely measurement-driven. 

As shown in Fig.~\ref{fig:tput_comparison}, the Release~16 scheme frequently suffers from sharp throughput degradation due to delayed beam updates as the UE moves through the dense urban environment. In contrast, UE-initiated beam reporting allows the UE to trigger a report as soon as beam quality degrades, enabling faster beam switching and shortening beam misalignment intervals. Consequently, the Release~19 scheme mitigates the severe throughput drops observed in the Release~16 baseline, thereby demonstrating that dynamic UE-triggered reporting is effective for maintaining robust service quality in time-varying channel environments.


\begin{figure}[t]
    \centering    
    \includegraphics[width=.45\textwidth]{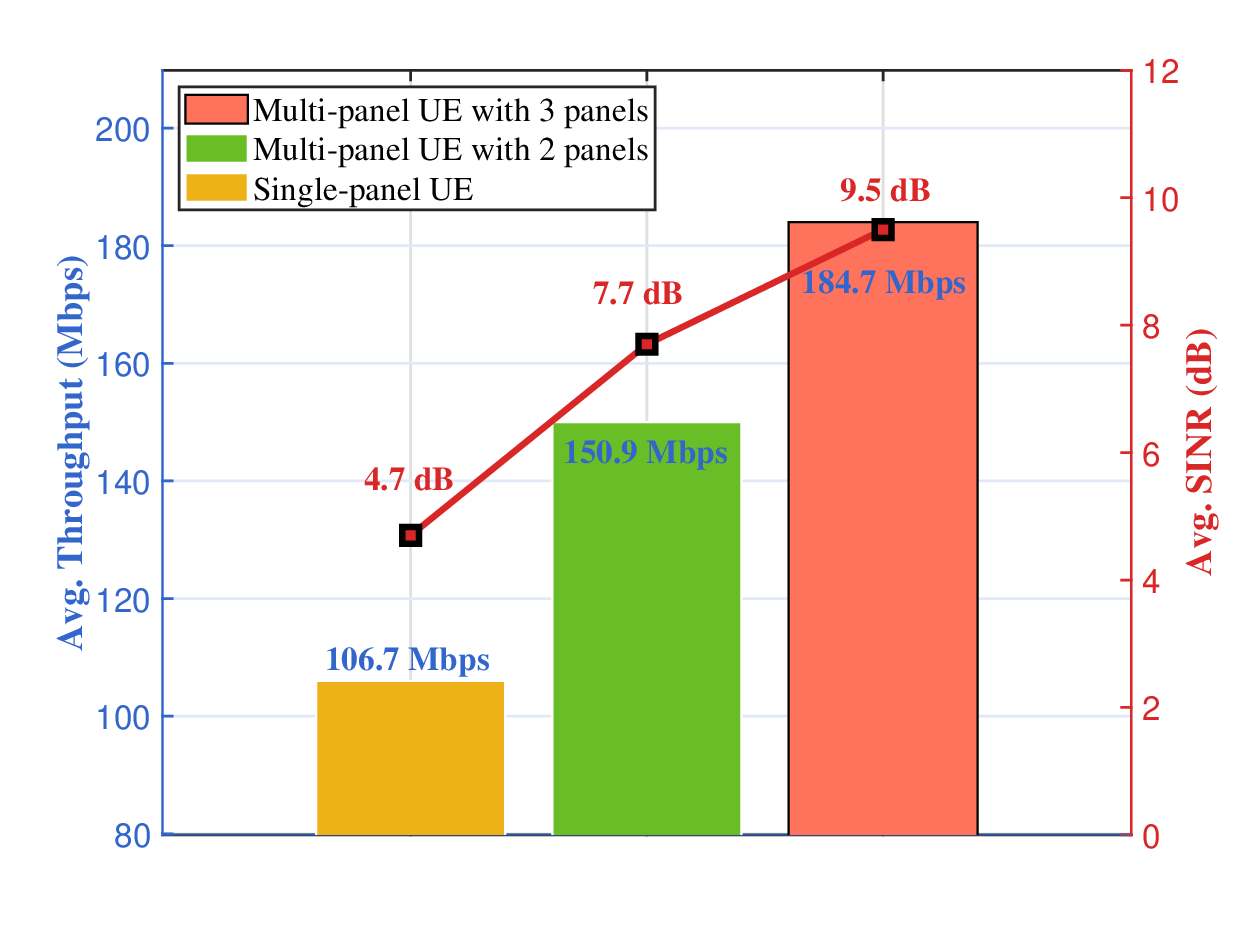}
    \caption{Average SINR and throughput comparison between single-panel and multi-panel UEs in a mobility scenario.}
    \label{fig:multi_panel}
\end{figure}

\subsection{Benefits of MPUE Architectures}
\label{Sec5-3}
Form-factor constraints and human-body-induced signal blockage are major UE-side factors that limit mMIMO performance. To validate the effectiveness of the proposed hardware-level approach, we compare a conventional single-panel UE with MPUEs equipped with two and three antenna panels.

Fig.~\ref{fig:multi_panel} shows a clear dependence of link robustness on the number of antenna panels. The average SINR of the single-panel UE is 4.7~dB, while increasing the number of panels to three improves the average SINR to 9.5~dB. This SINR improvement translates directly into higher data rates. As a result, the three-panel MPUE achieves an approximately 73.2~\% throughput gain relative to the single-panel UE, increasing from 106.7~Mbps to 184.7~Mbps. These results indicate that the MPUE architecture discussed in Section~\ref{Sec3-1} is not merely an optional design choice, but a critical hardware evolution for sustaining high-throughput communication under practical blockage conditions.

Beyond the quantitative gains, the simulation results provide two key insights. First, meeting the reliability and ultra-high data-rate requirements envisioned for 6G requires joint consideration of software-level evolution, such as UE-initiated reporting, and hardware-level innovation, exemplified by MPUE architectures. Second, the results highlight the value of digital twin platforms as a practical validation tool for designing and optimizing next-generation communication systems whose complexity continues to grow. Overall, these findings underscore that the UE should be treated not as a passive endpoint, but as a central intelligent node that can decisively shape system-level performance.


\section{Conclusions}
\label{Chap6}
This article presented an integrated analysis of the evolution of mMIMO systems from 5G to 6G, with a particular focus on the changing role of the UE. By examining standardization, practical implementation, and architectural innovation, we showed that the UE is transitioning from a measurement-and-feedback endpoint toward an active participant in next-generation mMIMO systems.
Through a review of 3GPP Releases~15 to 19, we summarized how UE capabilities have progressed from conventional CSI feedback toward more responsive measurement, reporting, and beam management. We also discussed practical device-level challenges, highlighting the importance of MPUE architectures, on-device processing, and energy-efficient baseband design. 
Overall, the success of 6G mMIMO systems requires coordinated UE--network co-design rather than network-centric innovation alone. Specifically, dynamic UE-triggered measurement and reporting can reduce reaction latency under fast channel dynamics, MPUE-based hardware evolution can mitigate form-factor and blockage limitations, and efficient on-device processing can enable scalable intelligence under tight power budgets. To foster continuous architectural innovation, remaining system-level challenges must be addressed, including reliable cross-device synchronization for UE cooperation, power-aware real-time inference in practical chipsets, and interoperable signaling support across heterogeneous deployment environments.

\section*{Acknowledgment}
This work was supported by Samsung Electronics Co., (IO251210-14274-01); This work was supported by Institute of Information \& communications Technology Planning \& Evaluation (IITP) under 6G·Cloud Research and Education Open Hub(IITP-2026-RS-2024-00428780) grant funded by the Korea government(MSIT); This work was partly supported by the Institute of Information \& Communications Technology Planning \& Evaluation(IITP)-ITRC(Information Technology Research Center) grant funded by the Korea government(MSIT)(IITP-2025-RS-2020-II201787, contribution rate 10~\%) 

\bibliographystyle{IEEEtran}
\bibliography{refs_all}

\section*{Biographies}
\vspace{-3em}
\begin{IEEEbiographynophoto}{Kwonyeol Park} (kwon10.park@kaist.ac.kr) received his M.S. degree in Electrical Engineering from Korea University in 2014. He is currently pursuing the Ph.D degree with the School of Electrical Engineering, KAIST. Since 2015, he has been with the S.LSI division, Samsung Electronics as a senior engineer.
\end{IEEEbiographynophoto}
\vspace{-2em}
\begin{IEEEbiographynophoto}{Hyuckjin Choi} (hyuckjin.choi@lge.com) received his Ph.D. degree in the School of Electrical Engineering, KAIST in 2024. He is now with the CTO division, LG Electronics as a senior engineer.
\end{IEEEbiographynophoto}
\vspace{-2em}
\begin{IEEEbiographynophoto}{Geonho Han} (ghhan6@etri.re.kr) received his Ph.D. degree in the School of Electrical Engineering, KAIST in 2025. He is now with the ETRI as a researcher.
\end{IEEEbiographynophoto}
\vspace{-2em}
\begin{IEEEbiographynophoto}{Gyoseung Lee} (iee4432@kaist.ac.kr) received his M.S. degree in Electrical Engineering from KAIST in 2023, where he is currently pursuing the Ph.D. degree.
\end{IEEEbiographynophoto}
\vspace{-2em}
\begin{IEEEbiographynophoto}{Yeonjoon Choi} (yeonjoon.choi@kaist.ac.kr) received his M.S. degree in Electrical and Computer Engineering from Sungkyunkwan University in 2018. He is currently pursuing the Ph.D. degree with the School of Electrical Engineering, KAIST. Since 2018, he has been with the S.LSI division, Samsung Electronics as a senior engineer.
\end{IEEEbiographynophoto}
\vspace{-2em}
\begin{IEEEbiographynophoto}{Sunwoo Park} (sunwoo0428@kaist.ac.kr) received his B.S. degree in Computer Science and Engineering from Hanyang University in 2018. He is currently pursuing the M.S degree with the School of Electrical Engineering, KAIST. Since 2018, he has been with the S.LSI division, Samsung Electronics as an engineer.
\end{IEEEbiographynophoto}
\vspace{-2em}
\begin{IEEEbiographynophoto}{Junil Choi}
(junil@kaist.ac.kr) received his Ph.D. degree in electrical and computer engineering from Purdue University in 2015. He is now with the School of Electrical Engineering at KAIST as an associate professor.
\end{IEEEbiographynophoto}

\vfill
	
\end{document}